\documentclass[lettersize,journal]{IEEEtran}
\usepackage{amsmath,amsfonts}
\usepackage{algorithmic}
\usepackage{algorithm}
\usepackage{array}
\usepackage[caption=false,font=normalsize,labelfont=sf,textfont=sf]{subfig}
\usepackage{textcomp}
\usepackage{stfloats}
\usepackage{url}
\usepackage{verbatim}
\usepackage{graphicx}
\usepackage{cite}
\usepackage[most]{tcolorbox} 
\usepackage{tablefootnote}  
\usepackage{booktabs}  
\usepackage{makecell}  
\usepackage{threeparttable}
\usepackage{multirow}
\usepackage{pifont} 
\newtcolorbox{rqanswer}{
	enhanced,
	breakable,            
	colback=gray!15,       
	colframe=black,      
	boxrule=0.4pt,       
	arc=3pt,             
	left=5pt, right=5pt, top=5pt, bottom=5pt 
}

\begin{document}
	
	\title{EstRTL: Functional Estimation Guided\\ RTL Code Generation}
	
	\author{Qi Xiong, Renzhi Chen, Bowei Wang, Yuqing Xiong, Libo Huang, and Lei Wang
		\thanks{Qi Xiong and Bowei Wang are with the College of Computer Science and Technology, National University of Defense Technology, Changsha, China (xiongqi@nudt.edu.cn).} 
		\thanks{Renzhi Chen is with the Defense Innovation Institute, Academy of Military Science (AMS), Beijing, China (chenrenzhi@qiyuanlab.com).} 
		\thanks{Yuqing Xiong is with the School of Computer Science and Technology, Shandong University, Qingdao, China.} 
		\thanks{Libo Huang is with the Key Laboratory of Advanced Microprocessor Chips and Systems, Changsha, China, and College of Computer Science and Technology, National University of Defense Technology, Changsha, China.}
		\thanks{Lei Wang is with the Defense Innovation Institute, AMS, Beijing, China and Qiyuan Lab, Beijing, China (wangleia@qiyuanlab.com).}}

\markboth{}{}


\maketitle
\begingroup
\renewcommand\thefootnote{}
\footnotetext{This work has been submitted to the IEEE for possible publication. 
	Copyright may be transferred without notice, after which this version may no longer be accessible.}
\endgroup

\begin{abstract}
	Optimizing register transfer level (RTL) code is of vital importance in hardware design. Large language models (LLMs) provide new methods for the automatic generation and optimization of RTL code, offering the potential to significantly accelerate the design process and reduce human effort. However, existing methods for generating RTL code often focus on model fine-tuning and the use of various expansion techniques to enhance the RTL code generation capabilities, lacking attention to the functional correctness. Ensuring that the generated RTL code not only compiles successfully but also behaves as intended in real hardware implementations remains a critical challenge. 
	
	To address this issue, we propose EstRTL, an LLM-powered collaborative agent framework for RTL code generation based on static functional score estimation. EstRTL operates a three-stage paradigm: Generation, Estimation and Correction. During the stages, the functional estimation agent statically evaluates the generated code based on score and assessment results, and decides whether to output the code directly, return it for regeneration, or forward it to the code correction agent. This framework can be applied to various LLMs that designed for RTL code generation, further enhancing the correctness of the generated code. By providing quantitative scores and human-readable requirements comparisons, it improves the transparency of AI-assisted RTL code generation. 
	
	Experiments show that EstRTL significantly improves the correctness of RTL code generation by generic LLM by 3.2\%-9.0\%, demonstrating the practical value of our system. The codes and experimental results are open-sourced at link: https://anonymous.4open.science/status/EstRTL-E200/.
\end{abstract}

\begin{IEEEkeywords}
	RTL Code Generation, Large Language Model,  Functional Estimation.
\end{IEEEkeywords}

\section{Introduction}

\IEEEPARstart{M}{anual} RTL design remains a time-consuming and error-prone process, especially for complex systems like modern processors where codebases can exceed hundreds of thousands of lines. For example, the construction of a high-performance RISC-V CPU using the widely recognized Chisel language requires approximately 60,000 lines of code\cite{Towards}. Designers must meticulously describe cycle-accurate behavior and optimize for performance and area, making the process both time-consuming and prone to human error. In recent years, language models (LLMs) have demonstrated outstanding performance in natural language processing (NLP) tasks\cite{b1}, enabling a wide range of applications such as text generation\cite{wu2024large}, translation\cite{elshin2024general}, and question-answering\cite{ehsan2025automatic}. Previous studies\cite{2023swe,2024swe,l2mac,codeagent} show that LLM can produce syntactically correct code from natural language prompts, reducing development time and improving productivity. This success has naturally extended interest from general-purpose programming to domain-specific languages. In particular, the application of LLMs to hardware description languages (HDLs), such as Verilog and VHDL, has attracted considerable attention\cite{multipl,hardfails,ahmad2024hardware,laeufer2024rtl}.

Currently, LLM-assisted RTL code generation primarily involves three directions: constructing high-quality datasets for domain-specific fine-tuning\cite{VeriGen,RTLCoder,VeriCoder,OriGen,betterv,AutoVCoder,codev,craftrtl,scalertl}, developing advanced augmentation techniques such as Retrieval-Augmented Generation (RAG) and Graph-Enhanced methods to enhance generation\cite{RTL++,RTLFixer}, and exploring novel paradigms for code generation and correction\cite{b19,Mage,b21}. Among these, research on using LLMs to generate and fix RTL code demonstrates broader applicability, as its methods can also be integrated into the first two approaches to further improve the correctness of RTL code. While the generation of RTL code is essential, the subsequent repair phase plays a crucial role in addressing potential issues that may arise during the initial generation. RTLFixer\cite{RTLFixer} leverages ReAct prompting to frame RTL error correction as an autonomous agent task. Spec2RTL-Agent\cite{spec2rtl} utilizes a multi-agent framework to automate the translation from complex natural language specifications into syntactically correct and functionally relevant Verilog code. However, current methods encounter challenges that the correctness verification of RTL code relies heavily on testbenches. Manual creation of testbenches is time-consuming and labor-intensive, while automatically generated testbenches by LLMs may lack comprehensiveness and accuracy\cite{qiu2024autobench,qiu2025correctbench}. Therefore, most existing methods solely focus on syntactic correctness rather than functionality. Without code evaluation results, LLM-generated and LLM-fixed RTL code often cannot be seamlessly integrated, thus are usually studied separately\cite{qiu2024autobench, bhandari2024llm}. Furthermore, most existing code fixing studies rely on binary pass/fail evaluation, which offers no intermediate feedback and thus lacks fine-grained guidance for iterative model optimization.

\begin{figure}[htbp]
	\centering
	\includegraphics[width=\linewidth]{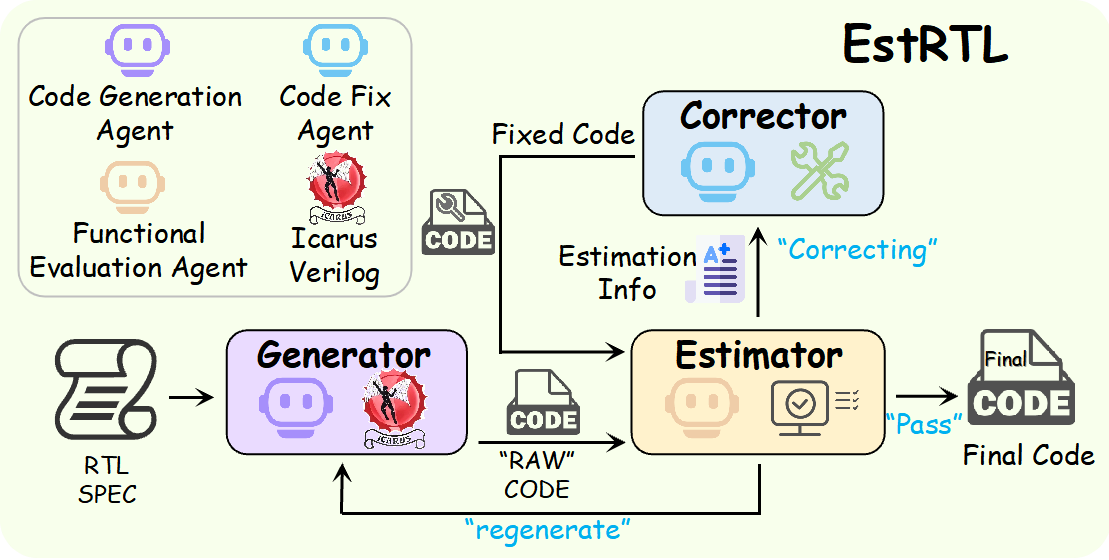}
	\caption{The overview of EstRTL. EstRTL operates a three-stage paradigm of Generation, Estimation and Correction.}
	\label{fig:1}
\end{figure}

In this work, we introduce EstRTL, an LLM-powered collaborative agent framework for RTL code generation based on static functional score estimation. As shown in Figure~\ref{fig:1}, EstRTL operates a three-stage paradigm: \texttt{Generation}, \texttt{Estimation} and \texttt{Correction}. Firstly, the RTL code specification is fed into the code generation agent, which produces code that is iteratively refined until it passes a syntax check. Then, the functional estimation agent performs static functional score estimation, generating both a score and an assessment result to judge the code quality. If the estimation indicates that the code quality is unacceptably low, it is returned to the code generation agent for regeneration; otherwise, if the estimation suggests that the code fails but still has potential, it proceeds to the code correction agent in the third stage, where the code is corrected and re-estimated.

We have verified this framework on the VerilogEval dataset. It leads to an improvement of 3.2\% to 9\% in RTL code generation accuracy across most of the sota commercial LLMs, such as Qwen-Plus, Qwen3-32B, DeepSeek-Chat, and GPT-4o. In addition to this, our experiments further validated the framework's effectiveness. Specifically, the functional estimation agent was able to distinguish between functionally correct and incorrect RTL code with precision exceeding 78\% across four commercial LLMs. The code correction agent demonstrated its capability by successfully fixing over 12\% of erroneous RTL code. Moreover, Our approach, when integrated with fine-tuned RTL code generation models like OriGen, RTLCoder, and CodeV within the EstRTL framework, achieved improvements in code generation and correction accuracy ranging from 3.3\% to 7.7\%. Static functional score estimation enhances the transparency of AI-assisted RTL code generation by providing quantified scores, offering intermediate feedback and fine-grained guidance for code correction. By estimating code functional correctness rather than labor-intensive testbenches, it bridges code generation and correction. 

Our contributions are summarized as follows:
\begin{itemize}
	\item A framework is developed for generating RTL code based on LLM-Agent, increasing the success rate of the generic LLM on the VerilogEval dataset by 3.2\%-9\%.
	
	\item We developed a method for estimating the functionality of RTL code without testbenches, achieving an average precision of 84.2\%. This approach greatly reduces the time and labor costs typically associated with traditional testbench creation, while still enabling effective functional evaluation of RTL code.
	
	\item Based on the code estimation, we developed an LLM-agent–based feedback loop for RTL code generation and correction, seamlessly integrating these three processes and fixing over 12\% of errors.
	
	\item We have open-sourced the EstRTL framework, which can be applied to various LLMs that designed for RTL code generation, further enhancing the correctness of the generated code.
\end{itemize}

\section{Background}

Current research has explored various methods for automatically generating RTL code using large language models\cite{VeriGen,RTLCoder,OriGen,betterv,AutoVCoder,codev,craftrtl,RTLFixer,Mage,promptv,scalertl,VeriCoder,RTL++,b19,b21}. Here, we briefly review the current mainstream methods for generating RTL code using LLMs and the technology of reverse requirement generation.

\subsection{LLMs for RTL Code Generation}

With the rise of LLM in the field of programming, more and more researchers are considering using LLM to automatically generate RTL designs from natural language descriptions. The automatic generation of RTL designs using Large Language Models (LLMs) can be broadly categorized into following several groups based on their core methodologies.

One line of work focuses on fine-tuning LLMs with domain-specific data. A foundational approach involves curating hardware-specific datasets to fine-tune general-purpose LLMs, tailoring them for the Verilog domain. Early efforts like VeriGen\cite{VeriGen} is fine-tuned on a large dataset constructed from Verilog files collected from GitHub and textbooks, while RTLCoder\cite{RTLCoder} extracts keywords and features from RTL code to generate Verilog code. Subsequent work has focused on improving the quality and correctness of the training data. For instance, CraftRTL\cite{craftrtl} constructs "correct-by-construction" data from formal specifications like Karnaugh maps and state-transition diagrams, while OriGen\cite{OriGen} and BetterV\cite{betterv} leverage code-to-code enhancement and generative discriminators, respectively, to create refined training examples that align with synthesis and verification goals. These efforts emphasize the critical role of high-quality, domain-specific datasets in enabling accurate RTL code generation.

To overcome the inherent knowledge limitations of static models, researchers have integrated external knowledge via advanced prompting and augmentation. Retrieval-Augmented Generation (RAG) is a prominent technique, used by AutoVCoder\cite{AutoVCoder} to fetch relevant code snippets from a hardware database and by RTLFixer\cite{RTLFixer} to correct syntax errors. Beyond RAG, other methods like RTL++\cite{RTL++} incorporate graph-enhanced representations to capture structural hardware information, providing a richer context for the LLM than natural language alone. These extension techniques enhance semantic understanding and improve the generated RTL quality.

A third direction involves Generation and Correction Paradigms. The ultimate goal of these efforts is to reliably generate and fix functional RTL. This has led to diverse sampling and agent-based strategies. MAGE\cite{Mage} employs multi-candidate sampling and selection to generate high-quality code variants, while RTLFixer\cite{RTLFixer} frames error correction as an autonomous agent task using ReAct prompting. A recent shift explores collaborative frameworks, such as the multi-agent approach of PromptV\cite{promptv}, which distributes the design tasks among specialized LLM agents. Conversely, CodeV\cite{codev} addresses a complementary task by using LLMs for code summarization, generating descriptions from existing RTL. The recent Spec2RTL-Agent\cite{spec2rtl} framework employs a structured multi-agent team to automate the translation from complex natural language specifications to syntactically correct and functionally relevant Verilog code. These methods reflect the growing trend of using multi-agent frameworks to address the complexity of RTL design and correction.

Despite the progress made in LLM-assisted RTL generation, several challenges still persist. Many existing methods focus on syntax, facing challenges in ensuring functional correctness, synthesis-aware optimization, and handling complex design constraints. Additionally, another challenge is verification and synthesis alignment. While approaches like VeriGen\cite{VeriGen} and BetterV\cite{betterv} attempt to address this, aligning the generated code with synthesis and verification objectives remains an area requiring further refinement.

\subsection{LLMs for Degugging and Functional Estimation}

One promising approach to improving the accuracy of RTL code generation and correcting errors is to integrate error-checking mechanisms into the code generation pipeline. This integration would enhance the reliability of the generated code and allow for automatic correction of mistakes once detected. In software engineering, reverse requirement generation and traceability methods have been widely explored to improve system understanding, code restructuring, and quality assurance.

In software engineering, various reverse engineering techniques have been proposed to bridge the gap between code and its original requirements. For example, reverse semantic traceability methods have been developed to link code artifacts back to their requirements, thereby enhancing consistency and ensuring that the system meets its intended specifications \cite{b13}. Feature-based reverse engineering approaches, which leverage feature models and graphical slicing techniques, have also been proposed to recover and reconstruct system specifications from the code \cite{b11}. Moreover, unified frameworks that jointly express code and requirements in a manner that bridges domain-specific knowledge with machine-readable attributes have shown great potential in improving system understanding and supporting software quality assurance \cite{b12}.

\begin{figure}[htbp]
	\centering
	\includegraphics[width=\linewidth]{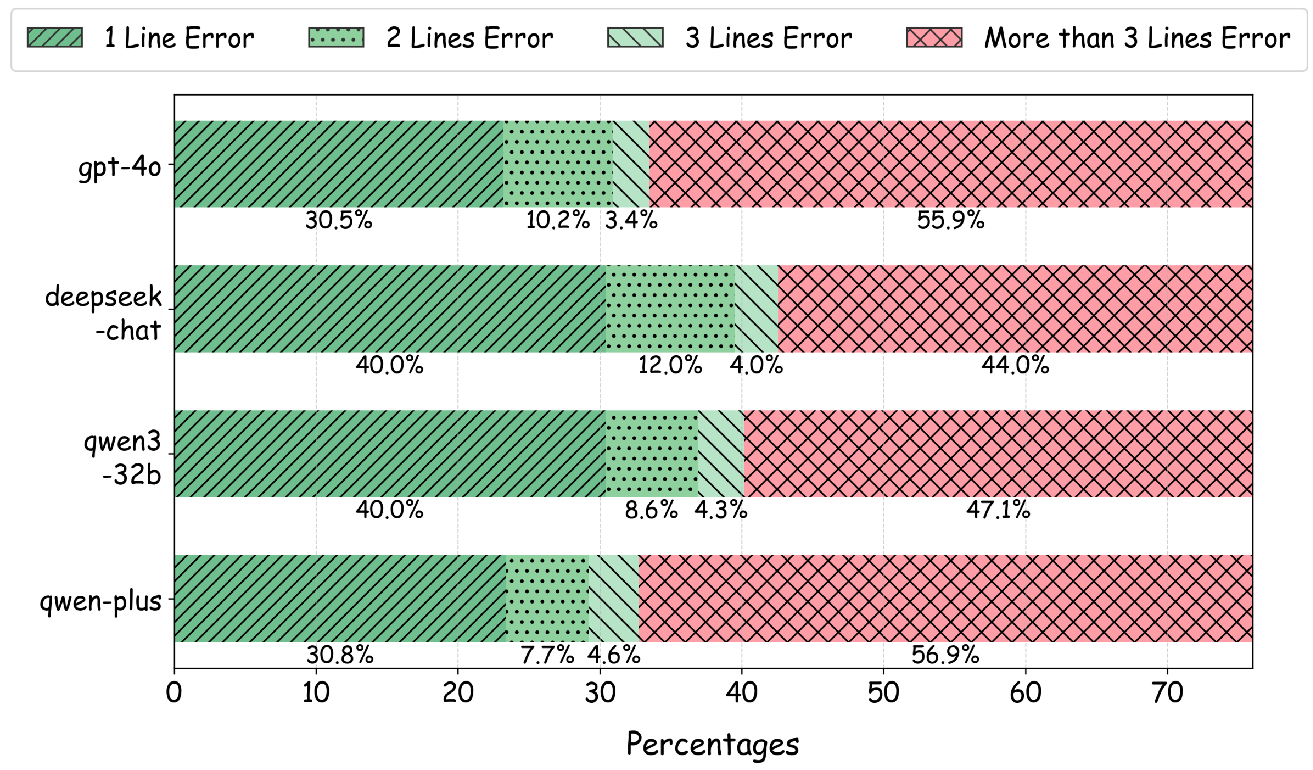}
	\caption{Distribution of errors in generated RTL code across DeepSeek-Chat, GPT-4o, Qwen3-32B, and Qwen-Plus. The chart categorizes errors based on severity, with horizontal bars representing the number of errors in each task. Red bars indicate tasks with more than three error lines, while green bars (ranging from dark to light) correspond to tasks with one, two, and three error lines, respectively.}
	\label{fig:8}
\end{figure}

More recently, the use of Large Language Models (LLMs) has gained traction for reverse engineering tasks. LLMs have been employed to reconstruct the original requirements from generated code, enabling a quantitative evaluation of the generated code's alignment with the requirements \cite{b15}. Additionally, natural language representations have been explored as intermediaries in the debugging process, offering a more flexible and intuitive approach compared to traditional code-level debugging methods \cite{NL-Debugging}. These advancements have demonstrated the efficacy of using LLMs in understanding and correcting software code, further motivating their application in hardware design.

Despite reverse requirement generation has been widely explored in software engineering, its application to hardware RTL code generation remains underdeveloped. However, this method holds great potential for integrating code generation and error correction, indicating that similar approaches could extend to hardware code generation, validation, and debugging.

\section{Motivation}

Through our experiments, we made several key observations. First, we found that the RTL code generated by LLMs often contains errors that are obvious and minor. Second, our experiments revealed that many of these common errors can be effectively addressed with relatively simple interventions. Specifically, we observed a correlation between the semantic quality of the generated code and its functional correctness. Here, we present our motivation experiments of this paper.

	
	

\subsection{What errors do LLMs make in RTL generation?}


Understanding the errors produced by LLMs is essential for improving the reliability of AI-assisted RTL code generation. To this end, we investigate a fundamental question: What errors do LLMs make in RTL generation? We task four general-purpose LLMs—DeepSeek-Chat, GPT-4o, Qwen3-32B, and Qwen-Plus—with generating RTL code for 156 RTL design tasks from VerilogEval-v2, and subsequently analyze each generated code instance at the task level.

As shown in Figure~\ref{fig:8}, the results of the experiment revealed that average 49\% of the errors in the generated RTL code were obvious and highly improbable to occur for a human hardware engineer. The most common errors were minor and typically involved only one or two lines of incorrect code. The chart illustrates the distribution of errors in the generated code across four LLMs. The horizontal bars are color-coded to represent different levels of error severity: red bars indicate tasks with more than three error lines, while green bars, from dark to light, represent correspond to tasks with one, two, and three error lines, respectively. Tasks with exactly one error line account for 30.5\%, 40\%, 40\%, and 30.8\% of the total tasks, highlighting the LLMs' struggle with seemingly simple yet critical design details.

\begin{figure*} 
	\centering
	\includegraphics[width=0.98\textwidth]{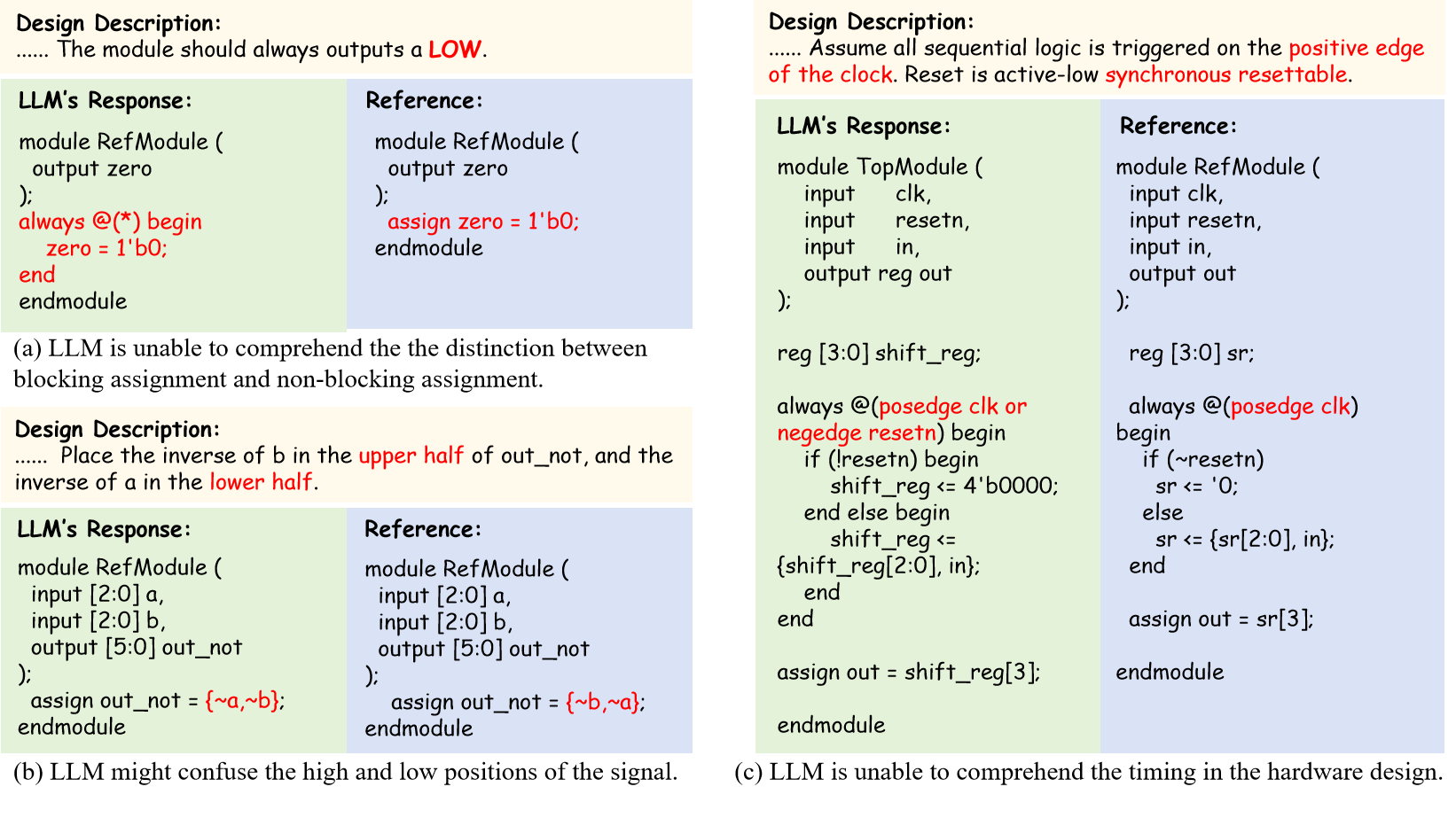}
	\caption{The figure presents the comparison between the generated RTL code and the reference code. The experiment found that LLM is unable to understand blocking assignments and non-blocking assignments, the high and low bits of signals, and sequential logic, etc.}
	\label{fig:9}
\end{figure*}

Figure~\ref{fig:9} presents several representative examples of errors in RTL code generated by LLMs. For instance, they fail to recognize the distinction between blocking assignment and non-blocking assignment. Take the example of a design where the module output must always be at a low level (The module should always output a LOW). An engineer would use a simple "assign out = 1'b0 ", whereas LLM might assign the value to the out signal using the always block. While both statements appear to achieve the same result, the latter introduces an always block, which typically incurs more resource consumption and unnecessary timing complexity. The code generated by LLM may lead to unnecessary resource waste, especially when unnecessary timing logic structures are used in simple logic designs, increasing the complexity of the hardware. Furthermore, LLMs often exhibit errors in handling the distinction between high and low bits in RTL code. For instance, given the instruction: "Place the inverse of b in the upper half of out\_not, and the inverse of a in the lower half." Human designers would intuitively connect out\_not[5:3] with the inverted value of b, and out\_not[2:0] with the inverted value of a. In contrast, LLM may mistakenly assign the inverted values of a and b to the wrong bit positions. It cannot fully grasp the concept of "high" and "low" in RTL code, nor can it accurately identify which specific bits of the signal correspond to these terms.

\begin{rqanswer}
	\textbf{Observation:} A large portion of the errors in the generated RTL code are relatively minor and localized. In many cases, only one to three lines of code deviate from the intended hardware behavior, while the overall structure and logic of the design remain largely correct.
	
	\textbf{Insight:} \textit{ Given the localized and minor nature of these errors, they present strong opportunities for effective correction, making it valuable to develop methods that can identify and fix such issues in LLM-generated RTL code.}
\end{rqanswer}

\subsection{Can LLM fix its errors?}

While a significant portion of the generated errors seem fundamental, a key finding from our analysis is that a substantial subset of these mistakes are highly correctable with minimal guidance. Providing the LLM with a brief, targeted prompt that highlights the specific oversight often leads to an immediate and correct revision.

For instance, when an LLM incorrectly implements a constant low output using an always block, a simple prompt such as "Use a continuous assignment statement instead of a procedural block for this combinational logic" is frequently sufficient to guide the model toward generating the more efficient assign out = 1'b0; construct. Similarly, in the case of bit-position errors, a clarifying instruction like "Ensure the inverted value of b is assigned to the upper three bits [5:3] of out\_not, and the inverted value of a to the lower three bits [2:0]" directly rectifies the model's conceptual confusion, resulting in accurate code.

This correctability is significant because it suggests that the LLM possesses the necessary foundational knowledge but fails to contextually apply it.  
Rather than needing extensive prompts, simply pointing out an error or providing a a brief, high-level hint is often enough for the LLM to self-correct. For example, when the model incorrectly uses an always block for a constant low output, a general prompt like "In general circumstances, use continuous assignment for combinational logic instead of procedural blocks" is often enough to guide the model toward generating the correct RTL code. This ability to self-correct with minimal guidance highlights the potential for improving LLM-generated code without extensive manual intervention or retraining.


\begin{rqanswer}
	\textbf{Observation:} Many of the errors are obvious and minor, which can be efficiently corrected with minimal adjustments or by providing targeted guidance.
	
	\textbf{Insight:}\textit{LLMs may lack full understanding in hardware design  principles, but they still possess foundational knowledge for accurate RTL generation. By identifying or pointing out specific issues, LLMs can often correct the errors on their own or with minimal prompts, offering a quick and efficient way to refine the generated code without the need for extensive retraining or major modifications.}
\end{rqanswer}

\subsection{Code Semantics vs. Code Functionality}

To this end, we aim to explore the correlation between the functionality of the generated code and its semantic consistency with the original requirements.

In our experiments, we tasked four different LLMs—Qwen-Plus, Qwen3-32B, DeepSeek-Chat, and GPT-4o—to generate RTL code based on 156 tasks from the verilogEVAL-v2 dataset. We then used these models to reverse-generate the requirements from the RTL code and calculated the semantic 
similarity between the generated requirements and the original 
\begin{figure}[htbp]
	\centering
	\includegraphics[width=\linewidth]{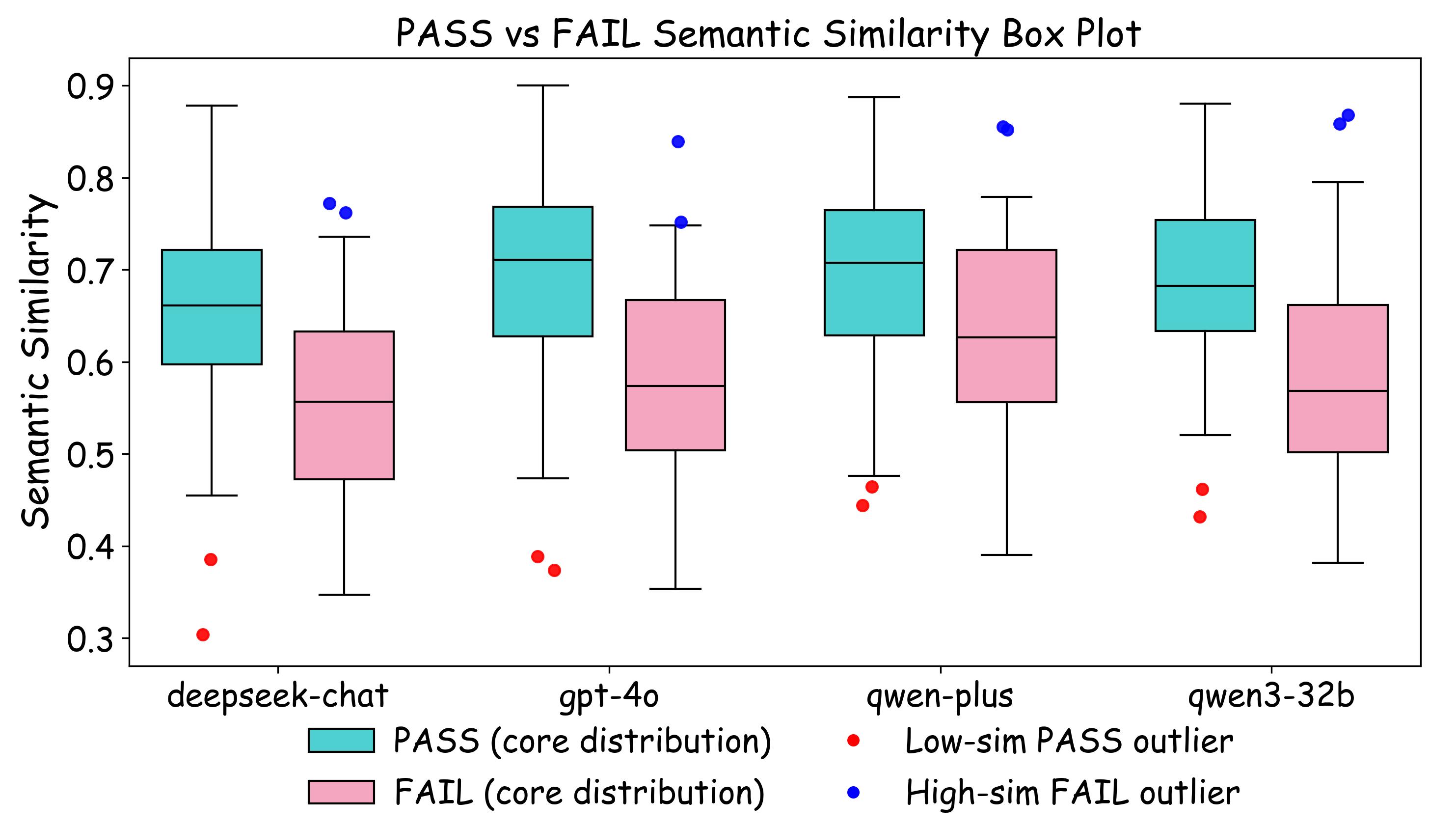}
	\caption{The results of comparing the average semantic similarity between the reverse generated requirements and the original SPEC for each generation task of the four large models, namely Qwen-Plus, Qwen3-32B, DeepSeek-Chat, and GPT-4o. PASS cases are shown in cyan, while FAIL cases are shown in pink. Red markers indicate unusually low-similarity PASS samples, and blue markers indicate unusually high-similarity FAIL samples. This indicates that there is certain correlation between the code semantics and its functionality. }
\label{fig:10}
\end{figure}
ones. 
Figure~\ref{fig:10} presents boxplots of 
\begin{figure}[htbp]
	\centering
	\includegraphics[width=\linewidth]{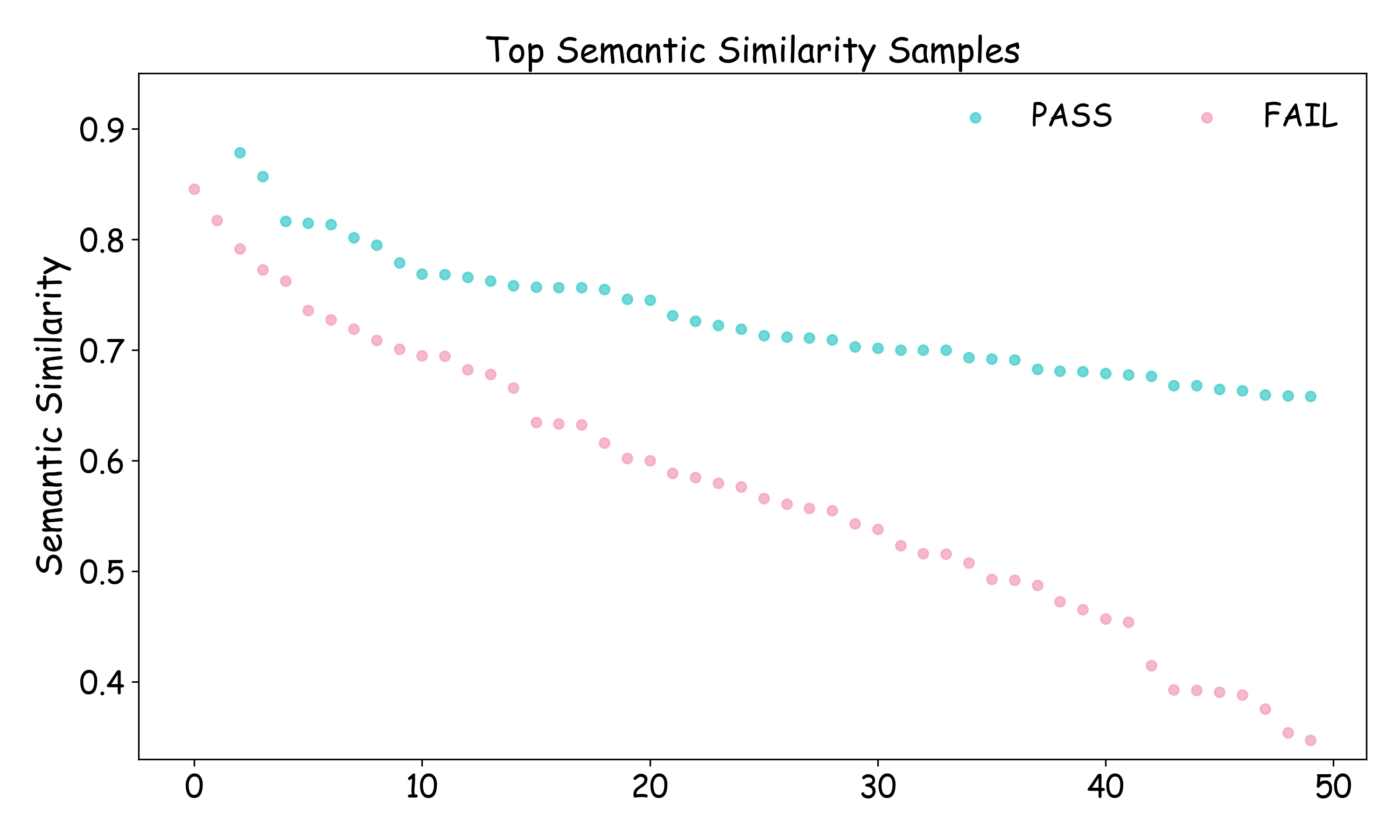}
	\caption{Top 50 semantic similarity scores for tasks generated by the DeepSeek-Chat model. Cyan points indicate tasks where the generated code passed verification (PASS), and pink points indicate tasks where the code failed verification (FAIL). The x-axis shows the sample sorted from highest to lowest similarity, and the y-axis represents the semantic similarity between the reversed and original requirements.}
	\label{fig:11}
\end{figure}
semantic similarity distributions for PASS and FAIL cases across four RTL code generation models. PASS cases are shown in cyan, while FAIL cases are shown in pink. In addition, red markers indicate unusually low-similarity PASS samples, and blue markers indicate unusually high-similarity FAIL samples, both identified as statistical outliers based on the interquartile range criterion. Across all evaluated models, the PASS distributions consistently exhibit higher median similarity and tighter interquartile ranges than the FAIL distributions. The limited overlap between the two boxplots, together with the scarcity of cross-class outliers, highlights a pronounced semantic separation between correct and incorrect RTL code. Low-similarity PASS samples and high-similarity FAIL samples appear only as sparse outliers. The boxplot reveals a clear separation between PASS and FAIL cases in terms of semantic similarity, indicating that semantic similarity provides a strong and consistent signal for functional correctness.



Figure~\ref{fig:11} shows the top 50 semantic similarity scores of the correct (PASS) and incorrect (FAIL) tasks of the generation tasks generated by the DeepSeek-Chat model. Tasks for which the generated code successfully passed functional verification are marked in cyan, while those that failed verification are marked in pink. It can be observed that the PASS tasks generally achieve higher similarity scores and exhibit a relatively smooth and gradual decrease from the highest to lower-ranked samples. In contrast, the FAIL tasks show lower similarity scores on average and a steep decline. This suggests that semantic similarity between the reversed and original requirements is correlated with the functionality of the generated code.


\begin{rqanswer}
\textbf{Observation:} The average semantic similarity between the reversed requirements and the original specifications is higher for tasks with correct RTL code, compared to tasks with incorrect code. 

\textbf{Insight:}\textit{ Code functionality is related to the semantic similarity between the original specifications and the reverse-generated requirements. By leveraging reverse requirement generation to assess semantic similarity, we can effectively gauge the quality and accuracy of generated RTL code.} 
\end{rqanswer}

\section{EstRTL:At a Glance}

EstRTL is a novel framework designed for LLM-based RTL code generation, aming to enhance the quality of RTL code generated by LLM through functional estimation. Figure~\ref{fig:1} illustrates the overall architecture of EstRTL, which integrates multiple collaborative agents to achieve end-to-end RTL development with enhanced correctness.

EstRTL takes natural language specifications and RTL code generation requirements as inputs and ultimately outputs optimized RTL code. It operates through a closed-loop workflow centered on three-stage paradigm: Generation, Estimation and Correction. Firstly, the RTL code specification is fed into the Code Generation Agent. If the code fail the Syntax checking of Icarus Verilog, the error message will be passed to the Code Generation Agent and the code will be regenerated until it pass the syntax check. This initial RTL code ("RAW Code") is then routed to the Estimator, which serves as the central decision-making hub of the framework. 

During this stage, the Functional Estimation Agent statically evaluates the generated code based on both a score and an assessment result to judge the code quality, determine the next workflow step: If the code quality is deemed "unacceptably low", it is returned to the Code Generation Agent for complete regeneration. If the code is assessed as "failing but improvable" it is forwarded to the Code Fix Agent for targeted correction.

The Code Fix Agent utilizes the assessment results to modify the code, producing a corrected version that is then recycled back to the Functional Evaluation Agent for re-estimation. This correction-re-estimation loop repeats until the code passes the functional check, at which point the final validated RTL code is output.

By separating code generation, functional estimation, and code correction into distinct stages, EstRTL ensures rigorous quality control at each step. This design not only ensures syntactic correctness but also significantly reduces the functional error rate prevalent in LLM-generated RTL code by: (1) filtering out low-quality code early via regeneration, (2) providing actionable feedback for corrections, and (3) enforcing iterative verification before finalization.

In the subsequent sections, we will elaborate on the phased decision-making mechanism of EstRTL, the static evaluation methods, and how the agent collaboration workflow achieves \ding{182} efficient iterative correction, \ding{183} functional correctness estimation, and \ding{184} seamless integration of generation and fixing workflows.

\subsection{Generation - Code Generation Agent}

\begin{figure}[htbp]
	\centering
	\includegraphics[width=\linewidth]{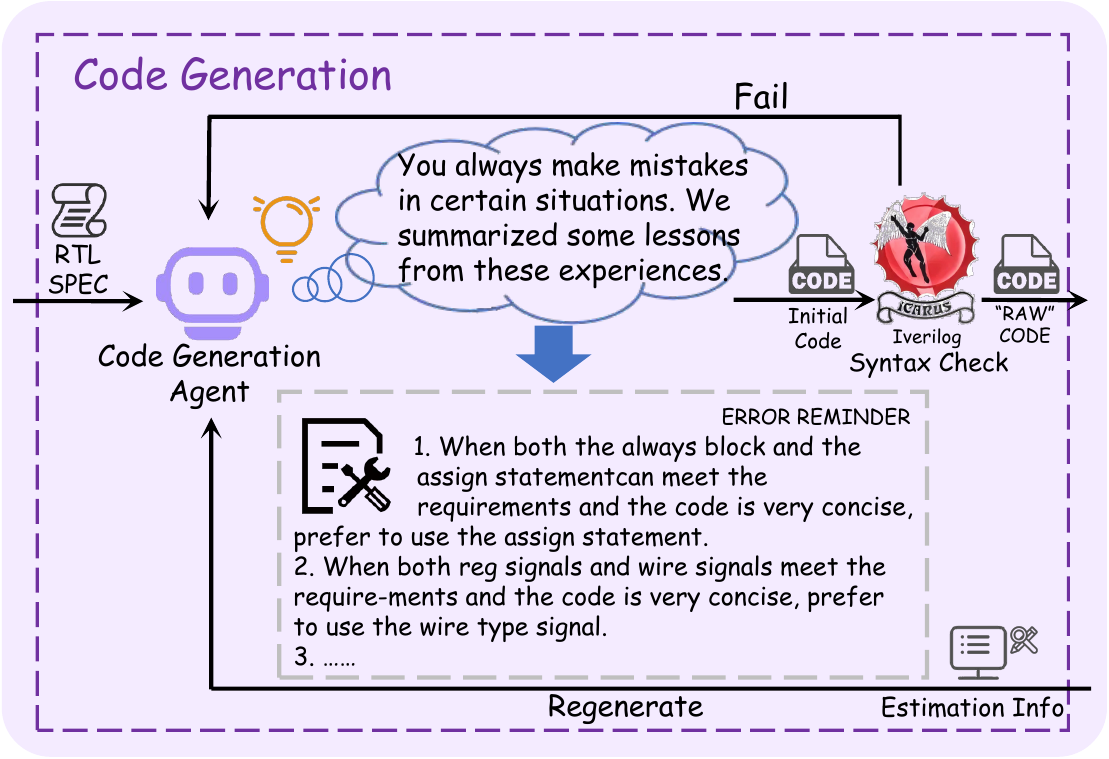}
	\caption{Framework of Code Generation. Code Generation process invovles Code Generation Agent to generate code ,Syntax-Checked and Estimation Info that guide regeneration.}
	\label{fig:2}
\end{figure}

As shown in Figure~\ref{fig:2}, 
the code generation part consists of a Code Generation Agent and a grammar checker. During the process of performing RTL code generation tasks, the results show that the Code Generation Agent would sometimes exhibit certain fixed errors in certain specific situations. For the errors that can be directly corrected, we summarize them as an error reminder and attach it to the prompt of the Code Generation Agent, guiding it to generate RTL code.

To improve grammatical correctness, the RTL code produced by the Code Generation Agent is passed to a grammar checker for parsing and verification against Verilog syntax rules. The checker identifies errors (e.g., missing port declarations, mismatched parentheses, or invalid operators) and returns the information to the agent for correction, iterating until the code passes. To ensure efficiency, the correction loop is limited to three iterations. If the code still fails, the system reports an error and skips to the next task to avoid infinite loops.

Based on our code estimation technology, the Functional Estimation Agent evaluates the quality of the generated RTL code by analyzing both its structural integrity and functional correctness. The agent provides a quantitative score as well as a qualitative assessment. If the code is judged to be of extremely low quality, the code is returned to the Code Generation Agent for complete regeneration, ensuring that subsequent refinement and correction efforts are built upon a more reliable baseline.

\subsection{Estimation - Functional Estimation Agent}

\begin{figure*}[htbp]
\centering
\includegraphics[width=0.999\textwidth]{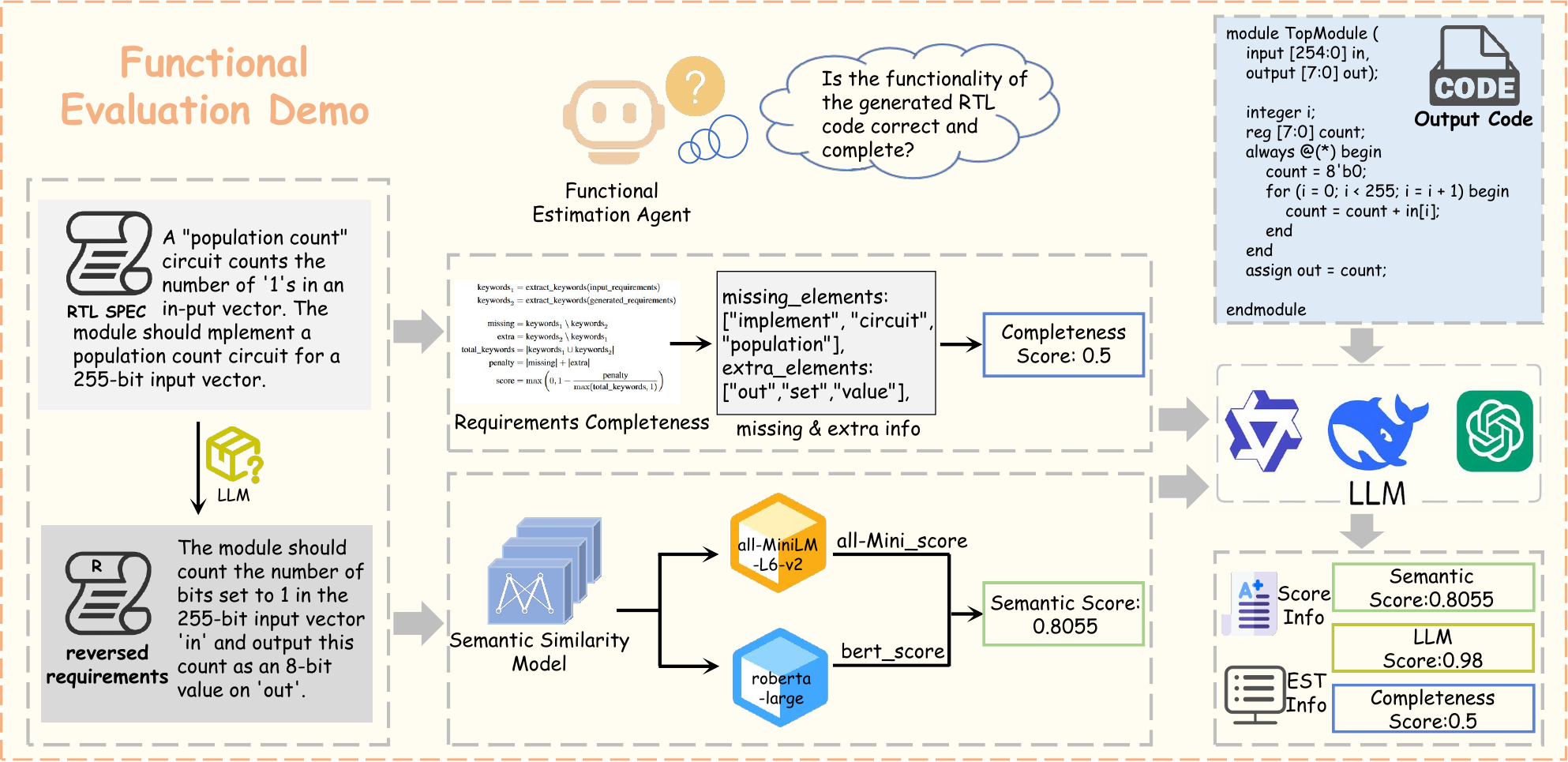}
\caption{A Demo of Functional Estimation part. The generated RTL code is reversed into requirements, compared with the originals for semantic similarity and completeness. Completeness and semantic similarity scores combined with the LLM comprehension score, is integrated into the final VFCS functional score.}
\label{fig:3}
\end{figure*}

We were inspired by the software engineering field and reference \cite{b15}, and designed the Functional Estimation Agent for analyzing the functionality of RTL code. Functional Estimation Agent mainly calculate three aspects of scores: semantic similarity score, completeness score, and LLM comprehension score. Firstly, the generated RTL code is input to the reverse requirement generator to reconstruct requirements. A pre-trained model then scores semantic similarity and completeness against the original requirements. These scores are provided to the LLM Judger to derive a comprehension score, and together they form the final VFCS (Functional Correctness) functional score.

The VFCS score is calculated using the following formula:

\begin{equation}
v_s = \sigma \left( S_{\text{token}} + S_{\text{sentence}} + C + L_{\text{LLM}} \right) 
\end{equation}
where \(v_s\) represents the VFCS score, \(S_{\text{token}}\) represents the token-level similarity score, \(S_{\text{sentence}}\) represents the sentence-level similarity score, \(C\) represents the completeness score, \(L_{\text{LLM}}\) represents the LLM score assigned by the LLM for the generated code. \(\sigma(z)\) is used for normalization.

\( S_{\text{token}} \) encodes the input text at the token level with the pre-trained language model roberta\-large, and calculates the bidirectional maximum cosine similarity between each token of the generated requirement and the reference requirement.
\(S_{\text{sentence}}\) uses the pre-trained model all-MiniLM-L6-v2 in the sentence Transformers library to encode the input requirement and the generated requirement at the sentence level. During the calculation process, the cosine similarity function in PyTorch is used to match the embedding vectors of the two sentences, thereby obtaining the sentence-level semantic similarity score.

\(C\) extracts key nouns, verbs, and appropriate nouns from the input, and calculates penalties based on missing or additional keywords. Assuming \( K_1 \) and \( K_2 \) represent the sets of keywords extracted from the input requirements and reversed requirements respectively, then \(C\) is calculated as the following function:

\begin{equation}
\begin{aligned}
C &= \max\!\Bigg(0,\, 1 - \frac{\phi}{\max\!\big(T_{total}, 1\big)}\Bigg).
\end{aligned}
\end{equation}
where \( \phi \) represents the penalty, calculated as the sum of the sizes of the missing (present in \( K_1 \) but not in \( K_2 \)) and additional (present in \( K_2 \) but not in \( K_1 \)) keyword sets, \(| K_1 \setminus K_2 |+|K_2 \setminus K_1|\). \( T_{total} \) represents the total number of keywords in both sets \( K_1 \) and \( K_2 \), which is given by \( |K_1 \cup K_2| \).

\(L_{\text{LLM}}\) takes original code requirements along with the code generated by the code generator as input, and supplements them with \( S_{\text{token}} \), \(S_{\text{sentence}}\), and \(C\). Based on this information, the LLM produces a comprehension score for the code.

Figure~\ref{fig:3} illustrates how the Functional Estimation Agent evaluates a “population count” circuit. The given RTL specification requires the module to count the number of ones in a 255-bit input vector and output the result as an 8-bit value. The Functional Estimation Agent first processes the RTL code using an LLM to generate a set of reversed requirements expressed in natural language. Then, semantic similarity models (such as all-MiniLM-L6-v2 and RoBERTa) are employed to compute embedding-based similarity between the reversed requirements and the original requirement text, yielding a semantic score (Semantic Score = 0.8055). In parallel, a requirement completeness check is performed to identify whether the reversed requirements covers the essential requirement elements. According to the predefined calculation formula, missing elements (e.g., “implement”, “circuit”, “population”) are counted as omissions, while extra elements (e.g., “out”, “set”, “value”) are flagged as extraneous. Based on this analysis, a completeness score (\(C\) = 0.5) is computed. Finally, the system integrates the semantic similarity score, completeness analysis, and an overall LLM-based evaluation (\(L_{\text{LLM}}\) = 0.98) to produce the final functional estimation. This example demonstrates that Functional Estimation Agent effectively combines semantic alignment and requirement coverage analysis to accurately assess functional correctness in a more comprehensive manner, enabling it to reliably distinguish high-quality implementations from low-quality samples.

\subsection{Correct - Code Correction Agent}

Inspired by Reference\cite{qiu2025correctbench}, we designed a Code Correction Agent based on Chain of Thought (COT). The Code Correction Agent operates during the conversation stage based on a large language model (LLM), using the model's reasoning ability to repair RTL code. The Code Correction Agent can access the following information: the generated RTL code, RTL design specifications, the original requirements, and the various scores in the scoring mechanism. When fixing the code, the Code Correction Agent gradually attribute the error content in the above-mentioned information, thereby achieving the correction of the code. The entire code repair process is divided into Reasoning and Correcting two stages. 

\begin{figure}[htbp]
	\centering
	\includegraphics[width=\linewidth]{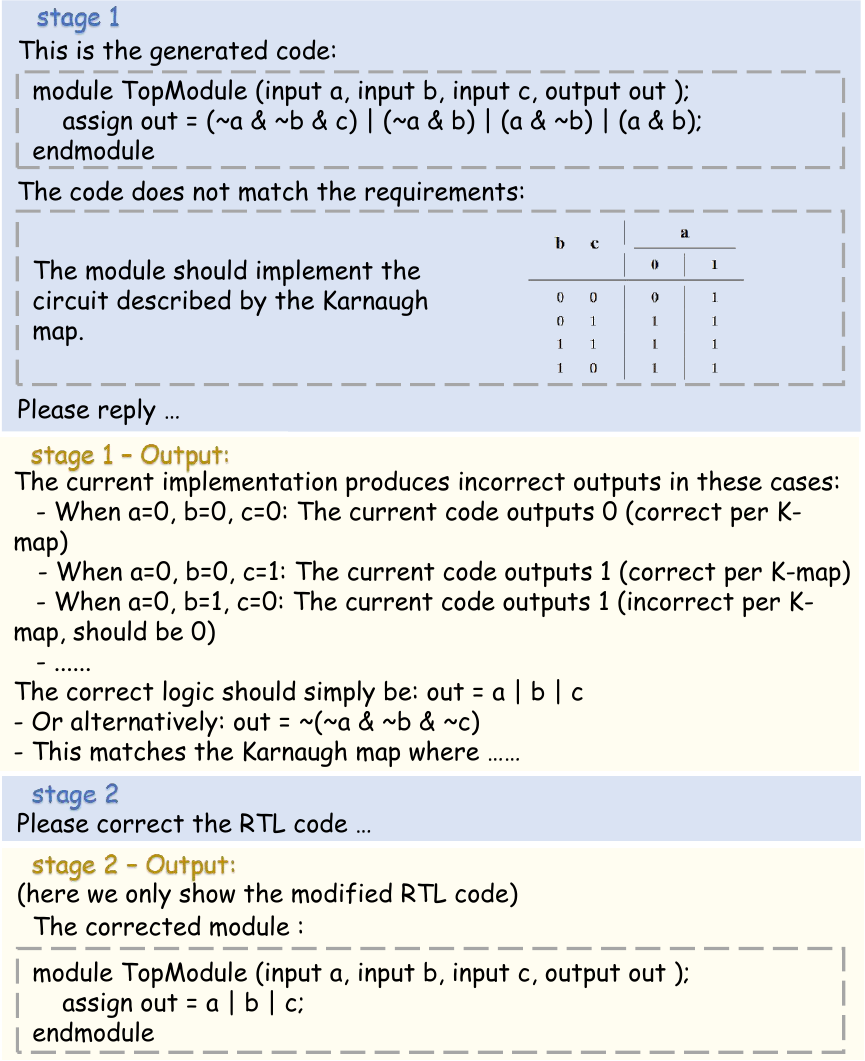}
	\caption{A Demo of Corrector. The RTL problem is the logic of signal out. Some details are omitted to save space.}
	\label{fig:4}
\end{figure}

Stage 1 - Reasoning: In this phase, the LLM is guided through a structured diagnostic process to analyze RTL code failures. The LLM first identifies which output signals exhibit errors and determines the specific input conditions that trigger these faulty behaviors and provide corresponding solutions in natural language (Which output signal has an error? Under what input signal conditions or set which data will this output signal have an error? How can this error be corrected?). This systematic approach helps pinpoint not just where errors occur in the code, but also why they manifest under particular operational scenarios. As shown in Figure~\ref{fig:4}, the first and second questions direct the LLM to combine insights about both the physical location of errors in the RTL code and their underlying logical causes, helping the model identify 
the problem. Subsequently, the LLM will propose a method based on natural language to solve these errors, based on the source and location of the RTL code error. Reasoning phase outputs a comprehensive error message that includes signal traces, triggering conditions, and proposed fixes.

Stage 2 - Correcting: In the second phase, building on the reasoning phase outputs, the LLM is guided to modify the RTL code. The process incorporates multiple safeguards, including RTL design specifications and error warnings to prevent recurrence of known issues. Each modification undergoes verification against the original failure conditions and preserve consistent coding style throughout all changes. In this way, the repair process not only focuses on correcting current errors but also proactively prevents similar issues from recurring, significantly enhancing the accuracy, stability, and reliability of the corrected RTL code.

This two-phase approach offers significant advantages over direct correction methods. By separating diagnostic reasoning from implementation, the framework forces comprehensive error analysis before any code modifications are attempted. It prevents the common pitfall of introducing new errors while fixing existing ones, which is a frequent occurrence when LLMs attempt direct corrections without proper contextual understanding. The reasoning stage builds the necessary contextual awareness by identifying not just error locations but also their triggering conditions and underlying causes, while the correction phase concentrates on precise technical implementation, significantly reducing the functional error rate in LLM-generated RTL. 

Example in figure~\ref{fig:4} illustrates the code correction process. In the first stage, the Code Correction Agent analyzes the generated RTL code against the given requirements. The analysis identifies mismatches between the implementation and the expected output, explains under which input conditions the code produces incorrect results, and highlights the erroneous logic. Based on this diagnosis, the Code Correction Agent provides natural language feedback, suggesting the corrected Boolean expression (e.g., out = a $|$ b $|$ c). In the second stage, the Code Correction Agent modifies the RTL code accordingly, replacing the faulty logic with the corrected expression and producing a functionally correct implementation. This example demonstrates how the Code correction Agent not only detects functional errors but also guides the code correction process in a structured manner.

\section{Experimental Setup}

\subsection{Research Questions}\label{AA}

Our experiment aims the following research questions:

RQ1: Can the RTL code generated by EstRTL be verified for correctness in terms of grammar and functionality? 


This question investigates the EstRTL framework’s ability to generate RTL code that is both grammatically correct and functionally accurate. Specifically, we assess EstRTL’s performance in terms of code quality, generation stability, and overall reliability. To answer this, we conduct comparative experiments using four distinct large language models, applying the dataset’s testbench for grammar and functional verification.

RQ2: What is the performance of the Functional Estimation Agent in assessing the functional correctness of RTL code? 


We explore the role of the Functional Estimation Agent in assessing RTL code functionality. By leveraging natural language requirements alongside the RTL code as input, the agent predicts the correctness of the code's functional behavior without human intervention. This question examines the accuracy and reliability of such automated assessments.

RQ3: In the EstRTL, how do the Code Correction Agent contribute to the outcomes? 


This question evaluates the contribution of the Code Correction Agent within EstRTL. Specifically, we analyze how this agent enhances code generation tasks, particularly in correcting errors in RTL code. Through quantitative analysis, we determine the extent of the agent’s contributions and present detailed data on its effectiveness in improving code quality.

RQ4: How are the external LLMs integrated into the framework of EstRTL?

This question examines whether EstRTL can be effectively integrated with specialized Verilog-focused LLMs. To demonstrate this, we incorporate four fine-tuned models—OriGen\cite{OriGen}, RTLCoder\cite{RTLCoder} and CodeV-QC\cite{codev}—all designed for Verilog RTL generation. 

RQ5: Does the Functional Estimation Agent exhibit robustness to different correctness thresholds?

This question explores whether the Functional Estimation Agent maintains consistent classification performance across varying threshold values, thereby demonstrating that the threshold selection is not sensitive and can be flexibly chosen within a broad range.

\subsection{Baselines}\label{AA}

To provide a comprehensive comparison, we evaluate several baselines that represent both traditional and advanced approaches to RTL generation:

Base models: These include single-pass RTL code generation approaches using both generic LLMs such as DeepSeek-Chat as well as RTL-specific models like CodeV\cite{codev}.

Agent-based systems: We include multi-agent frameworks that aim to enhance LLM-based RTL generation. These systems combine LLMs with specialized agents for code fixing and validation, including LLM + RTLFixer\cite{RTLFixer}, VeriAssistant\cite{b19}, OriGen\cite{OriGen}, and AutoVCoder\cite{AutoVCoder}.

Experimental models: For the experiments, we select four state-of-the-art general-purpose models, including Qwen-Plus, Qwen3-32B, DeepSeek-Chat, and GPT-4o. These models represent the latest available models under current conditions, with varying model sizes, enabling us to assess the performance across different scales of model capabilities in natural language processing and code generation for hardware design tasks.

\subsection{Dataset}

VerilogEval-v1-Human dataset\cite{2023verilogeval}: Released by NVIDIA in 2023, this dataset consists of 156 real-world hardware designs that vary in complexity. It contains human-written descriptions (from HDLBits) and machine-generated descriptions (via GPT-3.5). While machine-generated descriptions often suffer from verbosity and inconsistency with industrial practices, the human-generated set serves as the benchmark for evaluating RTL generation methods.

VerilogEval-v2 dataset\cite{b18}: In August 2024, the VerilogEval-v1 dataset was updated to include new tasks, specifically focusing on specification-to-RTL tasks in addition to code completion. This revision also introduces in-context learning examples and categorizes common errors in Iverilog. Notably, the MachineEval subset is no longer supported, and the dataset now offers more comprehensive test cases for model evaluation.
\subsection{Indicater}

To measure the effectiveness of the Functional Estimation in the EstRTL, we use the following evaluation metrics:

\textbf{TP (True Positives)}: the number of samples whose true labels are positive and are also predicted as positive.

\textbf{FP (False Positives)}: the number of samples whose true label is negative but are predicted as positive. 

\textbf{TN (True Negatives)}: the number of samples whose true labels are negative and are also predicted as negative. 

\textbf{FN (False Negatives)}: the number of samples whose true labels are positive but were predicted as negative. 

\textbf{Accuracy}: measures the overall correctness of a classification model, which is the proportion of samples that are correctly classified by the model out of the total samples. 

\textbf{Precision}: measures the proportion of samples predicted as positive by the model that are actually positive. 

\textbf{Recall}: measures the proportion of actual positive samples among the samples that are correctly predicted as positive by the model. Also known as the recall rate or sensitivity.

\textbf{Pass@k}: we follow the VerilogEval paper, using pass@k metrics to measure the RTL code generation accuracy. The pass@k metric estimates the proportion of code generation tasks that can be solved at least once in k attempts:
\begin{equation}
\textbf{Pass}@k = \mathbb{E}_{\text{problems}}\!\left[1 - \frac{\binom{n-c}{k}}{\binom{n}{k}}\right]
\end{equation}
where $n \ge k$ represents the total number of trials for each porblem, and \( c \) represents the number of trials that pass the functional check.

These metrics provide a comprehensive measure of the agent’s performance in accurately identifying correct RTL code and distinguishing it from incorrect or incomplete code.

\subsection{Parameter Setting}

\begin{table}[t]
\centering
\caption{parameters.}
\label{tab:models-hyper}
\setlength{\tabcolsep}{6pt}
\renewcommand{\arraystretch}{1.15}
\footnotesize
\begin{tabular*}{\columnwidth}{@{\extracolsep{\fill}} c | c c c}
\toprule
\textbf{Model} 
& \makecell{\textbf{Default} \\ \textbf{Temperature}} 
& \makecell{\textbf{Max} \\ \textbf{Tokens}} 
& \makecell{\textbf{Model} \\ \textbf{Size}}  \\
\midrule
\makecell{DeepSeek-Chat} 
& 1.0 
& 64K context 
& 671B (MoE)  \\
\midrule
\makecell{Qwen3-32B} 
& 0.6--0.7 
& 128K context 
& 32B  \\
\midrule
\makecell{Qwen-Plus} 
& 0.6--0.7 
& 128K context 
& 235B (MoE)  \\
\midrule
GPT-4o 
& 1.0 
& \makecell{128K context \\ 16K--32K output} 
& Proprietary  \\
\midrule
OriGen 
& N/A 
& N/A 
& 7B \\
\midrule
RTLCoder 
& N/A 
& N/A 
& 7B  \\
\midrule
CodeV 
& N/A 
& N/A 
& 7B  \\
\bottomrule
\end{tabular*}
\end{table}

In our experiments, we evaluate several large language models for RTL code generation, including Qwen-Plus, Qwen3-32B, DeepSeek-Chat, and GPT-4o, as well as task-specialized models such as OriGen, RTLCoder, and CodeV. For all models, we strictly follow their official inference configurations and do not introduce any additional manual parameter tuning. Parameters are shown in Table~\ref{tab:models-hyper}.

For correctness assessment, a Functional Estimation Agent is used to assign a continuous functional score to each generated RTL code by estimating its behavioral consistency with the task specification. Based on this score, we apply a fixed decision threshold of 0.54: generated RTL codes with scores greater than or equal to this threshold are classified as functionally correct, while those below the threshold are considered incorrect. This threshold is determined through experiments(RQ5) and remains fixed across all models and experiments.


\section{Experimental Results}

\subsection{RQ1: Feasibility and Effectiveness}

In the first study, we provided RTL code specification to the EstRTL and used the testbench accompanying 
the dataset to verify the generated RTL code. We selected four distinct LLMs as baseline models: Qwen-Plus, Qwen3-32b, DeepSeek-Chat, GPT-4o and compared them with other different models and methods. 
\begin{table}[t]
	\centering
	\caption{Main results}
\label{tab:main_results}
\setlength{\tabcolsep}{6pt}
\renewcommand{\arraystretch}{1.15}
\footnotesize
\begin{tabular}{c | c | c | c}
	\toprule
	\textbf{Method} & \textbf{LLM Model} &
	\textbf{\makecell{VerilogEval-\\v1-Human\\(Pass@1)}} &
	\textbf{\makecell{VerilogEval-\\v2\\(Pass@1)}} \\
	\midrule
	\multirow{4}{*}{Generic LLM}
	& Qwen-Plus &62.2   & 58.3 \\
	& Qwen3-32b   &45.5   & 51.3 \\
	& DeepSeek-Chat &73.1   & 65.4 \\
	& GPT-4o   &50.6   & 62.2 \\
	\midrule
	\multirow{3}{*}{\makecell{RTL-Specified\\LLM}}
	& RTLCoder\cite{RTLCoder} & 41.6 & 36.5 \\
	& ITERTL\cite{wu2025itertl}   & 42.9 & N/A  \\
	& CodeV\cite{codev}    & 53.2 & N/A  \\
	\midrule
	\multirow{2}{*}{\makecell{LLM + \\RTLFixer\cite{RTLFixer}}}
	& GPT-3.5 Turbo & 46.4 & 44.9 \\
	& GPT-4 Turbo   & 65.0 & 65.4 \\
	\midrule
	\multirow{3}{*}{VeriAssist\cite{b19}}
	& Claude-3 & 41.6 & N/A \\
	& GPT-3.5  & 34.4 & N/A \\
	& GPT-4    & 50.5 & N/A \\
	\midrule
	OriGen\cite{OriGen}
	& {\makecell{DeepSeek-Coder-\\7B + LoRA}} & 54.4 & 50.0 \\
	\midrule
	AutoVCoder\cite{AutoVCoder}
	& {\makecell{CodeQwen1.5-7B}} & 48.5 & N/A \\
	\midrule
	\multirow{4}{*}{EstRTL(Ours)}
	& Qwen-Plus &66.7 \textcolor{green}{(+4.5)}    & 63.5 \textcolor{green}{(+5.2)} \\
	& Qwen3-32b &52.6 \textcolor{green}{(+7.1)}   & 60.3 \textcolor{green}{\textbf{(+9.0)}} \\
	& \textbf{DeepSeek-Chat} & \textbf{76.3 }\textcolor{green}{(+3.2)} & \textbf{70.5 }\textcolor{green}{(+5.1)} \\
	& GPT-4o &58.3 \textcolor{green}{\textbf{(+7.7)}}     & 67.3 \textcolor{green}{(+5.1)}   \\
	\bottomrule
\end{tabular}
\begin{tablenotes}
	\footnotesize
	\item \textit{Note:} N/A denotes that evaluation on VerilogEval-v2 was not yet available when this paper was submitted, as the corresponding model is not publicly released.
\end{tablenotes}
\end{table}
We used the pass rate (the number of tasks verified by the test benchmark divided by the total number of code generation tasks) as the evaluation metric.

The results of functional correctness between our method and existing baselines are shown in Table~\ref{tab:main_results}. Our solution in DeepSeek-Chat achieves 76.3\% and 70.5\% accuracy in pass@1 evaluations on VerilogEval-v1-Human and VerilogEval-v2, respectively. Based on strong vanilla models such as GPT-4o, our method achieved significant improvements of 7.7\% on VerilogEval-v1-Human and 5.1\% on VerilogEval-v2. In addition, our approach outperforms various agent-based systems including LLM+RTLFixer, VeriAssist, and can be applied to other methods such as OriGen, AutoVCoder, and PromptV, further enhancing the performance of generating RTL code.

\begin{rqanswer}
\textbf{Answer to RQ1:} EstRTL can accurately generate RTL code according to the requirements. Compared with various other methods, the code generated by EstRTL is more correct, and it can further enhance the ability of various models to generate RTL code.
\end{rqanswer}

\subsection{RQ2: Functional Estimation Performance}

In the second study, we aimed to verify the effectiveness of the Functional Estimation Agent. We carried out multiple experiments on Qwen-Plus, Qwen3-32B, DeepSeek-Chat, and GPT-4o. We conducted a statistical analysis of the Functional Estimation Agent’s ability to separate functionally correct code samples from low-quality ones across different models, and quantified its actual performance.

Table~\ref{tab:classification_results} show that the Functional Estimation Agent is capable of effectively estimating the functional correctness of RTL code. Specifically, the precision in the deepseek-chat reaches 88.9\%, while the accuracy in the qwen3-32b reaches 78.2\%. Across models, the agent attains accuracies between 73.1\% and 78.2\% (mean 76.1\%), precisions between 78.7\% and 88.9\%, and recalls between 68.6\% and 76.8\%. These results indicate that the agent correctly distinguishes functional from low-quality samples in roughly three-quarters of all the cases.

\begin{table}[t]
	\centering
	\caption{The evaluation results of the Functional Estimation Agent on the RTL code generated by different models.} 
\label{tab:classification_results}
\setlength{\tabcolsep}{6pt}     
\renewcommand{\arraystretch}{1.15} 
\footnotesize
\begin{tabular}{c | c c c c | c c c}
	\toprule
	\multicolumn{1}{c|}{\textbf{LLM}} & \multicolumn{4}{c|}{\textbf{Confusion Counts}} & \multicolumn{3}{c}{\textbf{Evaluation Metrics}} \\
	\cmidrule(lr){2-5} \cmidrule(lr){6-8}
	& \textbf{TP} & \textbf{FP} & \textbf{TN} & \textbf{FN}
	& \textbf{Accuracy} & \textbf{Precision} & \textbf{Recall} \\
	\midrule
	\makecell{qwen-\\plus}     & 63 & 17 & 57 & 19 & 0.769 & 0.787 & 0.768 \\
	\midrule
	\makecell{qwen3-\\32b }    & 60 & 12 & 62 & 22 & 0.782 & 0.833 & 0.732 \\
	\midrule
	\makecell{deepseek-\\chat} & 72 &  9 & 42 & 33 & 0.731 & 0.889 & 0.686 \\
	\midrule
	gpt-4o        & 72 & 12 & 47 & 25 & 0.763 & 0.857 & 0.742 \\
	\bottomrule
\end{tabular}
\end{table}

\begin{figure}[htbp]
	\centering
	\includegraphics[width=\linewidth]{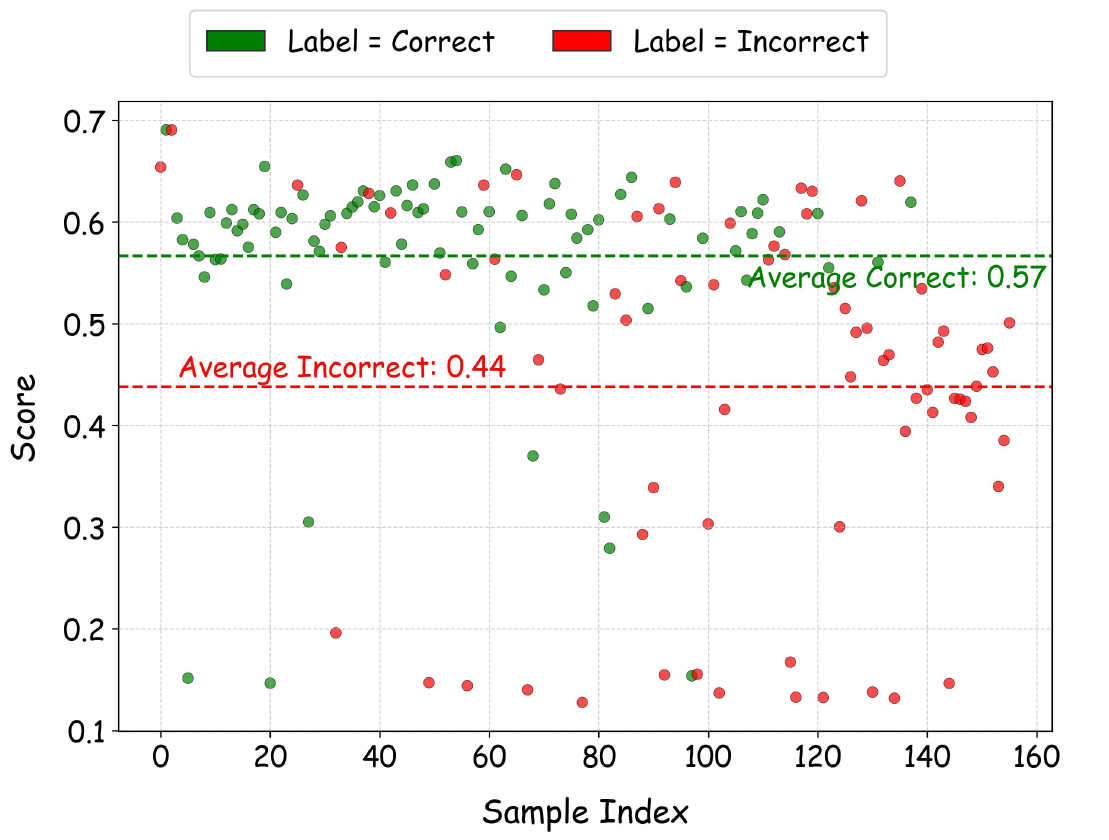}
	\caption{The performance of the Functional Estimation Agent in Qwen3-32b. The scores of the samples with correct functionality are generally higher, while the scores of the samples with incorrect functionality are lower. Function estimation results - Scatter plots in other experimental LLMs can be found in the uploaded open-source project.}
	\label{fig:6}
\end{figure}

Figure \ref{fig:6} illustrates the distribution of predicted scores across all samples, separated by their true labels. Green points represent functional correct samples, while red points represent incorrect ones. The green and red dashed horizontal line indicates the the average scores of correct and incorrect samples. The significant difference between the two dashed lines, showing the Functional Estimation Agent’s ability to discriminate between classes.

\begin{rqanswer}
\textbf{Answer to RQ2:} The Functional Estimation Agent can effectively estimate the functional correctness of RTL code. Compared with manual evaluation, it achieves reliable accuracy and precision across different models, enabling automated estimation of RTL code correctness directly from natural language requirements and generated code.
\end{rqanswer}

\subsection{RQ3: Code Correction Capability}

In this section, we isolate the Code Correction Agent and measure its ability to repair RTL code produced by different base LLMs. We manually selected samples of functional errors in RTL code, using the Code Correction Agent to correct them, then used corresponding testbench to test the corrected RTL. We report Correctly-Fixed, Incorrectly-Fixed, and the Correction Rate (Correctly-Fixed $-$ Incorrectly-Fixed / Sample-Size).

\begin{figure}[htbp]
	\centering
	\includegraphics[width=\linewidth]{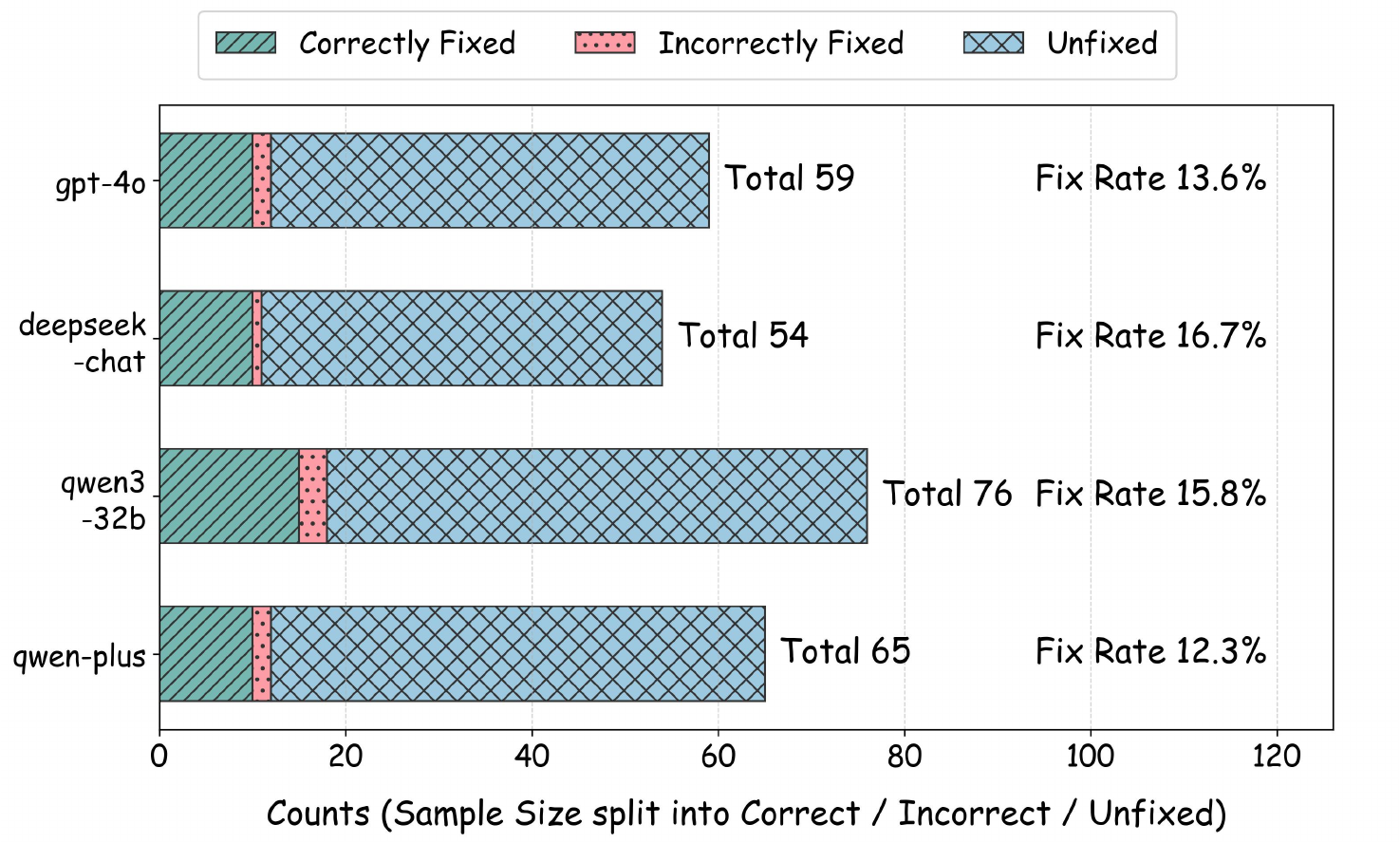}
	\caption{Classification evaluation of the Code Correction Agent. Bars show the proportions of correctly fixed, incorrectly fixed, and unfixed cases. Fix rates range from 12.3\% to 16.7\%, with DeepSeek-Chat achieving the highest rate.}
	\label{fig:5}
\end{figure}

As shown in figure~\ref{fig:5}, the agent delivers consistent, model-agnostic gains: on DeepSeek-Chat it achieves the highest net Correction Rate of 16.7\% (10 correct vs. 1 incorrect over 54 samples), followed by Qwen3-32B at 15.8\% (15 vs. 3 over 76). Across all models, the number of successful repairs (10–15) substantially exceeds the unfixed errors (1–3), demonstrating that the agent reliably converts failing cases into passing ones while introducing minimal new errors.

\begin{rqanswer}
\textbf{Answer to RQ3:} The Code Correction Agent makes a clear, quantifiable contribution to EstRTL. It provides a net absolute pass-rate uplift of 12–17\% on the evaluated error cases, generalizes across strong and mid-sized LLMs, and improves functional correctness with few regressions. This confirms that the code Correction agent has a certain effect in improving the quality of RTL code and increasing the pass rate of verification.
\end{rqanswer}

\subsection{RQ4: Model-Agnostic Extensibility}

In this study, we evaluate whether EstRTL can be effectively integrated with domain-specialized LLMs that have been fine-tuned specifically for RTL code generation. To this end, we substitute the LLM used in EstRTL’s code generation stage with four representative RTL-focused models—OriGen, RTLCoder,  and CodeV-QC—and evaluate the complete EstRTL pipeline, including generation, functional estimation, and correction. Experiments are conducted on both VerilogEval-v1-Human and VerilogEval-v2, using the dataset’s official testbenches for functional verification.

Table~\ref{tab:external_models} reports the results. Across all three specialized LLMs, integrating them into EstRTL consistently yields higher pass rates than their standalone generation performance. For instance, on VerilogEval-v1-Human, CodeV-QC exhibits the largest improvement of 6.9\% after integration, surpassing OriGen and RTLCoder, which improve by 3.3\% and 6.5\%, respectively. On VerilogEval-v2, CodeV-QC also achieves the highest gain of 11.5\%, while OriGen and RTLCoder also show substantial improvements, reaching 9.6\% and 7.7\%, respectively.. These improvements are achieved without altering the original decoding strategy or fine-tuning objectives of the external models, indicating that EstRTL functions as a plug-and-play framework rather than a competing modeling approach.

\begin{table}[t]
	\centering
	\caption{Integration of OriGen, RTLCoder, and CodeV-QC into EstRTL.}
	\label{tab:external_models}
	\setlength{\tabcolsep}{2.5pt}
	\renewcommand{\arraystretch}{1.2}
	\footnotesize
	
	\begin{tabular}{c|c|cc|cc}
		\toprule
		\multirow{2}{*}{\textbf{Method}} &
		\multirow{2}{*}{\textbf{\makecell{LLM\\ Model}}} &
		\multicolumn{2}{c|}{\textbf{\makecell{VerilogEval-v1\\-Human(Pass@1)}}} &
		\multicolumn{2}{c}{\textbf{\makecell{VerilogEval-v2\\(Pass@1)}}} \\
		\cline{3-6}
		& & \textbf{\makecell{Original}} & \textbf{\makecell{After \\Integrated}} & \textbf{\makecell{Original}} & \textbf{\makecell{After \\Integrated}} \\
		\midrule
		
		\multirow{1}{*}{OriGen\cite{OriGen}}
		& \makecell{DeepSeek-\\Coder-7B} & 54.4 & 57.7 \textcolor{green}{(+3.3)} & 50.0 & 59.6 \textcolor{green}{(+9.6)} \\
		\midrule
		
		\multirow{1}{*}{RTLCoder\cite{RTLCoder}}
		& \makecell{DeepSeek-\\coder-6.7B} & 41.6 & 48.1 \textcolor{green}{(+6.5)}& 36.5 & 44.2 \textcolor{green}{(+7.7)} \\
		\midrule
		
		\multirow{1}{*}{CodeV-QC\cite{codev}}
		& \makecell{Qwen2.5-\\Coder-7B} & 53.2 & 60.1 \textcolor{green}{(+6.9)}& 44.2 & 55.7 \textcolor{green}{(+11.5)}\\
		
		
		\bottomrule
	\end{tabular}
	
\end{table}

Moreover, the resulting pass rates remain stable across models of different sizes and training pipelines, suggesting that EstRTL does not introduce incompatibilities with RTL-specialized architectures. The consistent performance uplift demonstrates that the EstRTL framework can generalize to LLMs beyond the base models used in earlier experiments, reinforcing the framework’s robustness and extensibility.

\begin{rqanswer}
\textbf{Answer to RQ4:} EstRTL integrates seamlessly with specialized RTL-focused LLMs. When the LLM in EstRTL’s code generation stage is replaced with OriGen, RTLCoder, or CodeV-QC, the framework consistently enhances functional correctness on both VerilogEval-v1-Human and VerilogEval-v2. These results show that EstRTL does not conflict with fine-tuned RTL models; instead, it provides a model-agnostic framework capable of boosting their performance in RTL code generation.
\end{rqanswer}

\subsection{RQ5: Parameter Robustness}

\begin{figure}[htbp]
	\centering
	\includegraphics[width=\linewidth]{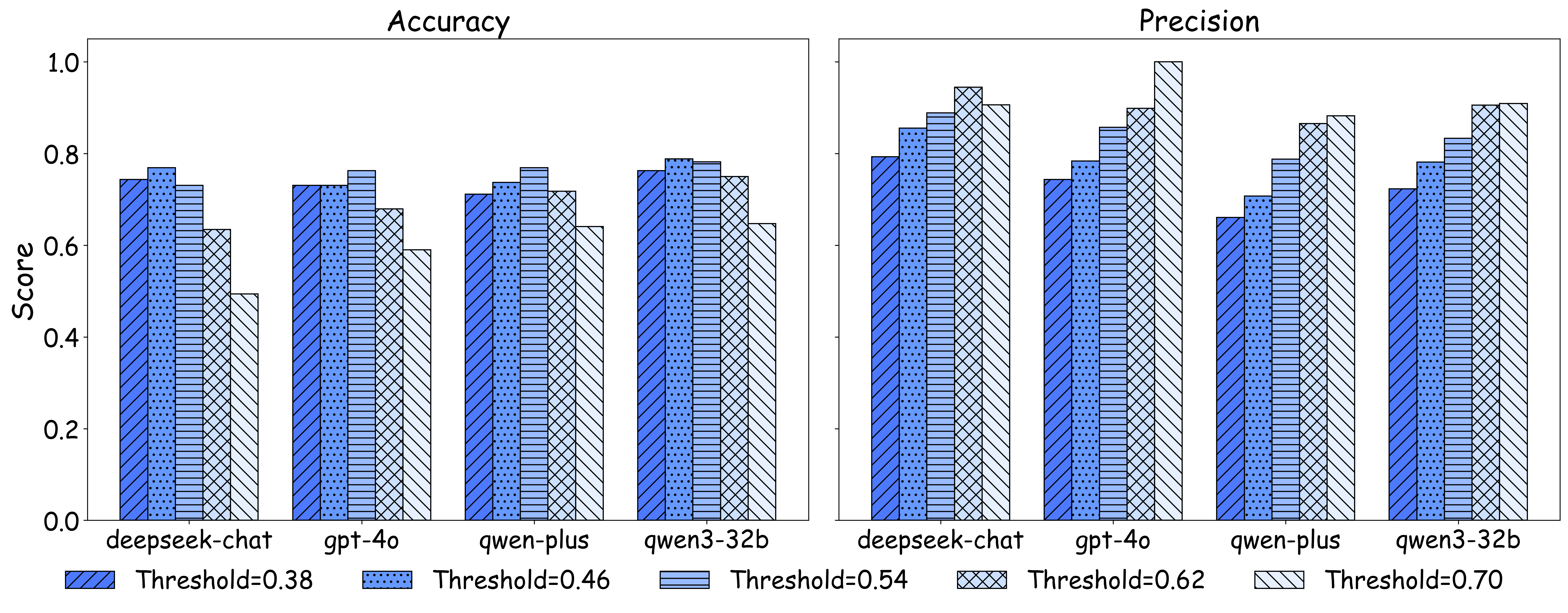}
	\caption{The left panel shows the accuracy and the right panel shows the precision of four large language models: DeepSeek-Chat, GPT-4o, Qwen-Plus, and Qwen3-32B. Each colored bar corresponds to a different threshold value used to convert the model’s final score into a PASS/FAIL prediction, as indicated in the shared legend below the plot. The figure highlights how varying the decision threshold impacts model evaluation metrics, allowing a visual comparison of robustness across LLMs.}
	\label{fig:12}
\end{figure}

To evaluate the robustness of EstRTL with respect to the decision threshold used by the Functional Estimation Agent, we examine the impact of threshold variation on overall performance. Specifically, we consider a range of threshold values and report the resulting accuracy and precision across four generic models: deepseek-chat, qwen-plus, qwen3-32b, and gpt-4o. The results are illustrated in Figure~\ref{fig:12}.

Across all evaluated models, accuracy generally increases as the threshold rises from 0.38 to around 0.54, after which it either plateaus or slightly decreases at higher thresholds. Precision exhibits a similar upward trend, as it is defined by 
$\mathrm{precision} = \frac{TP}{TP+FP}$, 
and typically the number of false positives (FP) decreases faster than the number of true positives (TP) when the threshold is raised, leading to higher precision. Despite these variations, the performance curves remain smooth without abrupt drops.

This stability holds consistently across models with different sizes and training paradigms, demonstrating that the robustness of the Functional Estimation Agent is not model-specific. These results suggest that the threshold parameter is easy to select in practice and does not require careful per-model calibration. As a result, EstRTL can be reliably deployed with a fixed threshold across diverse RTL code generation models without sacrificing performance.

\begin{rqanswer}
\textbf{Answer to RQ5:} EstRTL exhibits strong robustness to the decision threshold used by the Functional Estimation Agent. Varying the threshold within a practical range yields stable accuracy and precision across all evaluated models, indicating that EstRTL does not rely on finely tuned parameter settings. This stability enables the use of a fixed threshold across different RTL code generation models without sacrificing functional correctness.
\end{rqanswer}

\section{Conclusion}

In this work, we developed an innovative framework for generating RTL code based on LLM-Agent, significantly enhances the performance of generic LLMs when applied to the VerilogEval dataset, resulting in a notable increase in the success rate, ranging from 3.2\% to 9\%. Furthermore, our framework is fully open-source, which provides a valuable resource for the research community to further experiment, refine, and adapt the approach to a wide range of use cases and applications. This contribution not only advances the state of the art in RTL code generation but also emphasizes the critical importance of code functionality and error correction in ensuring the reliability and correctness of generated hardware descriptions. 
Our work provides a robust and adaptable prototype, serving as a foundation for future research and development in RTL code generation, as well as in broader areas of hardware design automation.

\section*{Acknowledgments}


This work was supported in part by the National Natural Science Foundation of China under Grants 62372461, 62032001, 62203457, and 62406335.
The authors used generative AI, specifically DeepSeek, solely to assist in polishing the grammar, wording, and syntax of this manuscript to improve readability. The authors reviewed and edited the content as needed and take full responsibility for the content of the publication.

\bibliographystyle{IEEEtran}  

\bibliography{references}

\vfill

\end{document}